# Design and Performance of the ARIANNA Hexagonal Radio Array Systems

S. W. Barwick[1], E. C. Berg[1], D. Z. Besson[2,3], E. Cheim[4], T. Duffin[1], J. C. Hanson[1,2], S. R. Klein[5], S. A. Kleinfelder[4*], T. Prakash[4], M. Piasecki[2], K. Ratzlaff[6], C. Reed[1], M. Roumi[4], A. Samanta[4], T. Stezelberger[5], J. Tatar[1,7], J. Walker[1], R. Young[6], and L. Zou[4]

[1]Dept. of Physics and Astronomy, University of California, Irvine, [2]Dept. of Physics and Astronomy, University of Kansas, [3]Moscow Physics and Engineering Institute, [4]Dept. of Electrical Engineering and Computer Science, University of California, Irvine, [5]Lawrence Berkeley National Laboratory, [6]Instrumentation Design Lab, University of Kansas, [7]Center for Experimental Nuclear Physics and Astrophysics, University of Washington

*Abstract*— We report on the development, installation and operation of the first three of seven stations deployed at the ARIANNA site's pilot Hexagonal Radio Array in Antarctica. The primary goal of the ARIANNA project is to observe ultra-high energy (>100 PeV) cosmogenic neutrino signatures using a large array of autonomous stations each dispersed 1 km apart on the surface of the Ross Ice Shelf. Sensing radio emissions of 100 MHz to 1 GHz, each station in the array contains RF antennas, amplifiers, 1.92 G-sample/s, 850 MHz bandwidth signal acquisition circuitry, pattern-matching trigger capabilities, an embedded CPU, 32 GB of solid-state data storage, and long-distance wireless and satellite communications. Power is provided by the sun and $LiFePO_4$ storage batteries, and the stations consume an average of 7W of power. Operation on solar power has resulted in ≥58% per calendar-year live-time. The station's pattern-trigger capabilities reduce the trigger rates to a few milli-Hertz with 4-sigma thresholds while retaining good stability and high efficiency for neutrino signals. The timing resolution of the station has been found to be 0.049 ps, RMS, and the angular precision of event reconstructions of signals bounced off of the sea-ice interface of the Ross Ice Shelf ranged from 0.14 to 0.17°. A new fully-synchronous 2+ G-sample/s, 1.5 GHz bandwidth 4-channel signal acquisition chip with deeper memory and flexible >600 MHz, <1 mV RMS sensitivity triggering has been designed and incorporated into a single-board data acquisition and control system that uses an average of only 1.7W of power. Along with updated amplifiers, these new systems are expected to be deployed during the 2014-2015 Austral summer to complete the Hexagonal Radio Array.

## I. INTRODUCTION

THE ARIANNA project (Antarctic Ross Ice-shelf ANtenna Neutrino Array) is a surface array of radio receivers planned to span ~1,000 $km^2$ of the Ross Ice Shelf in Antarctica, viewing ~0.5 Teratons of ice [1-4]. The project will detect radio waves originating from high energy neutrino interactions with atoms in the ice via the Askaryan Effect [5].

Neutrino interactions produce a shower of secondary particles, plus, for $\nu_\mu$ charged current interactions, an energetic muon. The secondary particles produce an electromagnetic or hadronic shower which extends over a length of many meters increasing with energy, with a transverse dimension of a few centimeters. For wavelengths much larger than this transverse dimension, electromagnetic radiation is coherent, so depends on the net charge in the shower. Compton scattering of atomic electrons, and annihilation of shower positrons on atomic electrons both contribute to lead to a net negative charge in the shower, leading to an intense Cherenkov radiation pulse, with a peak electric field that scales linearly with the shower energy. The frequency range of the radiation depends on the angle of observation of the shower. In ice, near the Cherenkov angle of about 56 degrees, the coherent radiation extends up to a maximum frequency of about 1 GHz; away from the Cherenkov angle, the cutoff is lower in frequency. ARIANNA is designed to improve the sensitivity to neutrinos with energies in excess of $10^{17}$ eV by at least an order of magnitude relative to existing limits [6, 7]. ARIANNA's goals includes a confirmation and measurement of the Greisen-Zatsepin-Kuzmin neutrino flux [8, 9], which results from cosmic rays interacting with the diffuse cosmic microwave background, and to measure the neutrino-nucleon cross-section.

ARIANNA takes advantage of unique geophysical features of the Ross Ice Shelf [10, 11]. The water-ice interface of the ice shelf acts as a nearly-perfect mirror for radio pulses generated by extremely high-energy neutrinos traveling downward and interacting in the ice [12]. The ice's long attenuation length allows for the detection of direct and reflected radio pulses at the surface. This and the ice shelf's relative proximity to McMurdo Station (~100 km away) significantly simplifies the deployment of a large array. A ridge known as Minna Bluff separates the ARIANNA site from McMurdo Station, and by this and its uninhabited location, the site has been found to be essentially free of anthropogenic noise. Being a surface array, ARIANNA stations are easy to deploy, maintain and upgrade.

* Corresponding author: Stuart A. Kleinfelder, University of California, 4416 Engineering Hall, Irvine CA, 92697, U.S.A. Voice: 949-824-9430, Email: stuartk@uci.edu.

Each ARIANNA station contains RF antennas, amplifiers, triggering, digitization, computing, power management, data storage, long-distance wireless networking and satellite communications, solar power and battery backup, plus experimental wind power. Four stations have been installed, including an early prototype deployed in 2011 [4] and the three HRA stations deployed in December of 2012, which are the subject of this paper. A block diagram of an HRA station configuration is shown in Fig. 1.

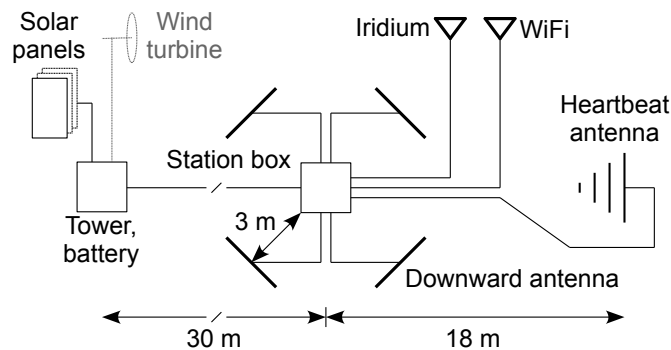

Fig. 1: Schematic station overview showing basic elements and distances (not to scale) of an ARIANNA Hexagonal Radio Array station.

ARIANNA is a surface array, with most components buried less than a meter beneath the snow surface. This design has many practical advantages when compared with deep-ice experiments [13-17], the most obvious of which is that no drilling is necessary to deploy the stations, saving an enormous amount of fuel, environmental impact, expense, effort and time. Surface deployment imposes fewer geometric constraints on receiver antennas and electronic systems than deep-ice designs. Cabling between antennas and electronics, etc., are minimal and at the surface. Deployed equipment is fully retrievable, and servicing or upgrades are eminently possible while keeping most of the installed infrastructure intact.

Sections II-VI of this paper focuses on the design of the major subsystems of the stations deployed in 2012 as part of a pilot phase of the ARIANNA project, known as the Hexagonal Radio Array. Section VII describes the system software for station monitoring and remote control. It also outlines the data collection, transmission and archiving procedures. Section VIII provides a discussion on the operational performance of the power systems, monitoring systems, trigger rates and environmental influences, and evaluation of the data quality. Section IX concludes with a discussion of improved data acquisition electronics intended for the completion of the HRA during the 2014-2015 Austral summer.

## II. THE ARIANNA HEXAGONAL RADIO

In 2010, the National Science Foundation approved a pilot program of the ARIANNA project, called the Hexagonal Radio Array (HRA), consisting of 7 stations dispersed on the ice in a hexagonal grid with 1 km between neighboring stations. The HRA's focus is to develop the technologies needed for a network of autonomous stations that achieve the performance necessary for the physics aims of the full-scale ARIANNA project. Stations must provide their own power, and must allow unattended remote monitoring, data retrieval and control. Operation at temperatures down to -30C or lower, and survival during harsh Antarctic conditions is a necessity. The electronics must be highly-sensitive over a 100-1,000 MHz frequency range and perform without themselves creating any radio frequency noise in this spectrum. Stations must be cost-effective and quick to deploy. Adapting to these requirements, and growing experience with instrument deployment in Antarctica's harsh conditions, has resulted in a rapid evolution of increasingly more robust, higher-performance, lower-powered and lower cost HRA hardware.

### A. HRA Timeline and Configuration

An early prototype station including a new 1.92 G-samples/s waveform acquisition and advanced real-time triggering system ("ATWD") was deployed in December of 2011 at Site D in Fig. 2 [4]. In December of 2012, ARIANNA deployed the first three HRA stations at sites A, C and G in Fig. 2, and converted the old Site D station to a weather monitoring post and WiFi repeater. This second-generation ARIANNA design replaced the prototype's hand-assembled electronics with a unified, mass-produced printed-circuit electronics system, replaced separate hand-constructed solar panel and wind turbine support structures with an integrated commercial tower system, and made many other refinements resulting in a much lower power (7 W average), much lower cost, lighter weight, lower noise, better calibrated and much faster and easier to deploy system. The station's reconstructed angular precision was found to be 0.17 degrees (see Section VIII-*G*)

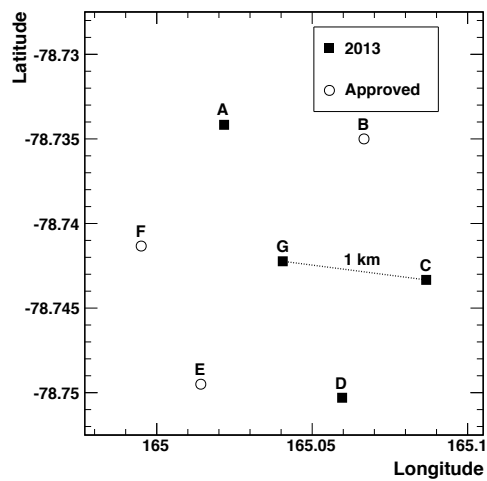

Fig. 2: Position and designation of the existing and approved HRA sites.

A curtailment of Antarctic operations during the 2013-2014 Austral summer permitted only a brief service mission. During this, the surface-array benefits of ARIANNA allowed a number of HRA system changes to be made, leading to performance that reached the levels needed for ARIANNA's physics goals. On-site radio reflection studies confirmed that the ARIANNA stations can achieve timing resolution of ~49 ps (see Section VIII-*F*).

The NSF has approved deployment of the HRA's remaining four stations during the 2014-2015 Austral summer. Plans include simplifications of the power tower, including the integration of the communications antennas. Improved amplifiers with flatter frequency response, improved stability and with integrated band-pass filtering and limiting have been fabricated. A simpler, lower-cost, lower-power, single board data acquisition system incorporating a new multi-channel signal acquisition chip, including deeper waveform storage and simplified trigger formation, will also be deployed.

### B. HRA System Overview

Each ARIANNA HRA station deployed thus far is divided into two major components: a power tower and an instrumentation and communications box with associated antennas. A power tower is seen in Fig. 3 (left), and a communication mast is seen (right) with an omni-directional antenna for mesh-connected wireless communications with McMurdo Station via a repeater on Mt. Discovery, plus an antenna for Iridium satellite short-burst messaging.

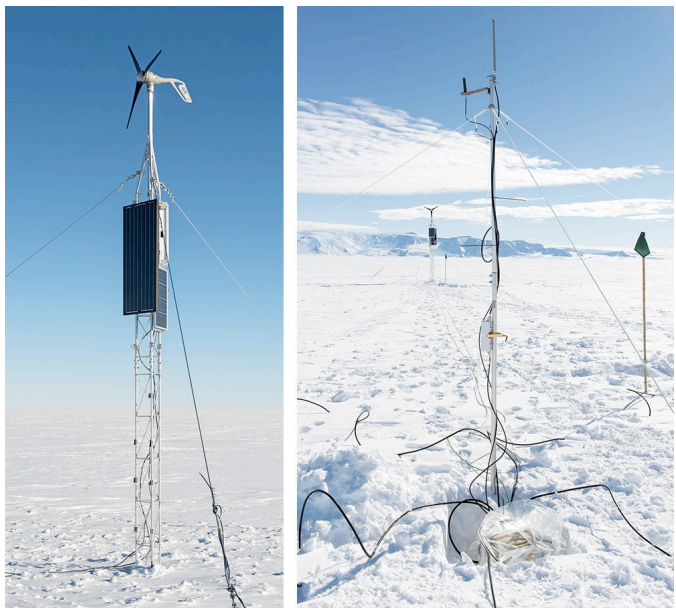

Fig. 3: An HRA system on the Ross Ice Shelf during deployment in 2012, showing a power tower (left), a communication's mast (right) with the tower in the background, a station box inside clear plastic at the foot of the mast (later buried), and a flag marking the location of one of four buried downward-pointing instrumentation antennas.

Section VII describes the power-tower, which includes 160W of solar panels, experimental inclusion of a 150W wind turbine, and an insulated battery box buried at the foot of the tower. For the 2012 deployment, the power and electronics assemblies were separated by about 30m due to concerns over potential RF noise emanating from the wind turbine. Hence, a separate communications mast was deployed along with the instrumentation box. The wind turbines were removed from the HRA systems during the 2013-2014 service mission due to reliability issues and to remove them from consideration as a source of anthropogenic noise. For the completion of the Hexagonal Array, no wind turbines will be deployed, and unified power, instrumentation and communications systems is planned (see Section IX).

A station and amplification box assembly (see Sections III and IV) is seen at the foot of the communications mast (Fig. 3, right), wrapped in plastic to prevent ice built-up on its connectors, etc., and was later buried. Four signal acquisition antennas are disbursed surrounding the communications mast, with the position of one seen marked by the green flag on the right-hand side of the photo.

### III. ANTENNAS AND AMPLIFICATION

#### A. Antennas.

Each station includes four log-periodic dipole antennas (Creative Design Co. model CLP5130-2N), positioned as two orthogonal pairs of parallel antennas 6 meters apart, pointing straight down into the ice. These 50 Ohm antennas have 17 elements and are about 1.4m long, with the span of its largest tines being 1.45m. The frequency response is quoted by the manufacturer as ranging from 105-1,300 MHz with a VSWR of 2.0:1 or less across the band (in air; in snow, their lower frequency limit is expected to be 70-80 MHz, e.g. in [1]). The forward gain is quoted as 11-13 dBi, with a front-to-back ratio of 15 dB. An example plot of the antenna's measured effective height (ratio of the induced voltage to the incident field) in the E-plane and H-plane in a common 40° off-axis angle is shown in Fig. 4 [18]. The antennas are connected via 5 meter LMR-400 cable (N-type connectors on both ends) to an RF-tight box containing four radio-frequency amplifiers (Fig. 5). Band-pass filtering leaves a frequency range of 100-1,000 MHz intact.

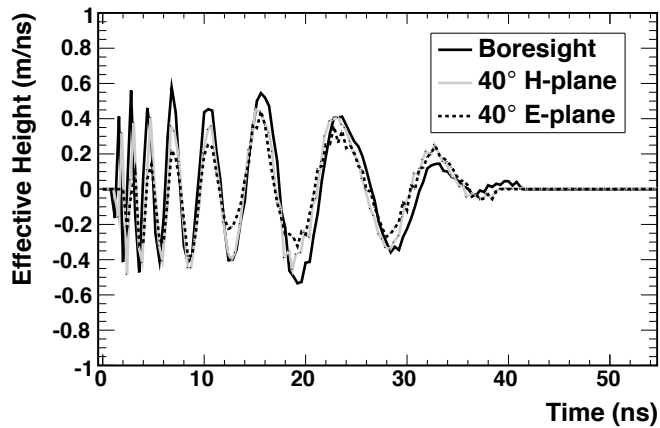

Fig. 4: Example antenna effective height vs. off-axis angle for the E and H-planes at 0° ("boresight") and 40° off-axis.

#### B. Amplification.

Each amplifier consists of four AC-coupled 1.5 GHz GaAs gain stages (Avago MGA-68563) with inter-stage filtering, yielding about 50-70 dB of gain over the frequency range of interest (Fig. 6). Power is conditioned in the main data acquisition enclosure and is supplied via coaxial cable to the amplifier box. Each amplifier consumes about 250 mW of power at 3.3V. Amplifiers are housed in individual brass enclosures that help prevent cross-talk between stages and between amplifiers. The amplifier's output range must be matched to the signal sampling and digitization electronics, and hence attenuation and limiting components were added to the amplifier's outputs. The limiting components cause compression of large signals, e.g., those above about half of the full 1.5V output range, as seen in Fig. 7.

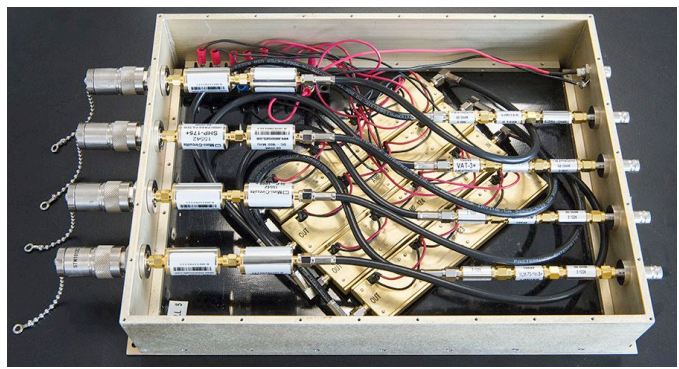

Fig. 5. The 2012 ARIANNA station's amplifier box (inputs to the left, outputs and power to the right). It contains four amplifiers, each with four AC-coupled 1.5 GHz GaAs amplifier stages with inter-stage bandwidth shaping. The box includes two RF filters per channel to constrain the frequency range to that of interest (~100-1,000 MHz), and output limiting and attenuation to optimize matching with the station's electronics.

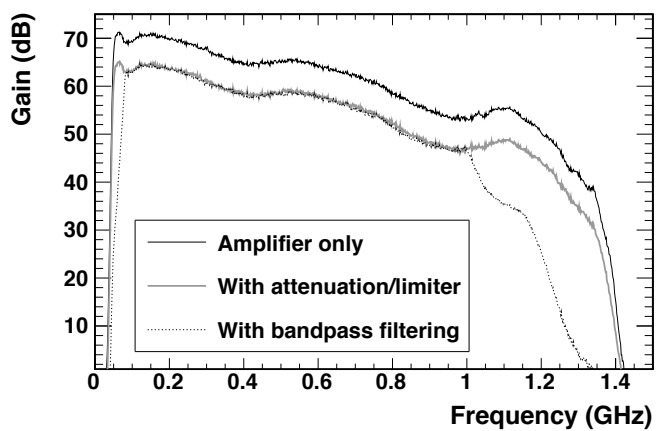

Fig. 6: ARIANNA amplifier gain vs. frequency plot. From top to bottom, the curves show the amplifier alone, the amplifier plus output attenuation and limiting, and the former with input band-pass filtering.

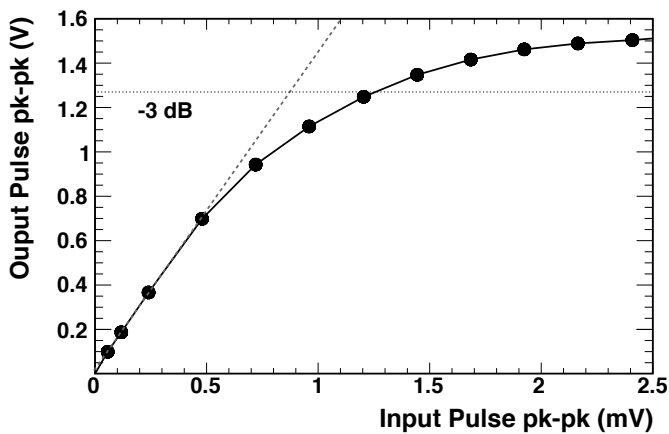

Fig. 7: Amplification system response to an impulsive input signal of varying magnitude, including input band-pass filtering and output limiting and attenuation. Note that the vertical scale is three orders of magnitude greater than the horizontal scale (V vs. mV).

### C. Heartbeat generation.

In order to monitor the health and stability of the ARIANNA stations, each station includes an auxiliary antenna that can transmit a radio-frequency signal to the other, instrumented, antennas. This "heartbeat" signal is typically triggered by the system software with some periodicity such as 1 Hz or less. The heartbeat pulse is produced by an FPGA on the system's motherboard, and its width is set in firmware to be about 1.5 ns full-width at half-maximum. The resulting signal is sent via LMR 600 cable to the same model log-periodic dipole array antenna as the receiving antennas. The heartbeat antenna is laid ~18 meters away from the center of the station with its E-plane parallel to the ground (i.e., flat on the surface), pointing back at the center of the station's antenna array, and aimed approximately along the diagonal between the four instrumentation antennas (i.e., each instrumentation antenna, pointing down, presents its face at a ~45 degree angle to the heartbeat antenna, as seen in Fig. 1). Monitoring the heartbeat pulse is a simple way to verify the functionality of virtually the entire signal acquisition, processing and transmission chain, and is also of potential use to watch for changes due to temperature, etc.

## IV. DATA ACQUISITION SYSTEM

A new HRA data acquisition system has been prepared and deployed. The main advances of this system are reduced power consumption, greatly improved manufacturability, lower cost, lower noise, improved physical integrity, lighter weight and more compact dimensions. The over-all power consumption has been reduced from ~30 W to ~7W during typical data taking, with as little as 0.6W possible in a minimum-power maintenance mode. This power reduction maximizes the control, communication and acquisition time on solar power and batteries during days of waxing and waning sun and/or heavy overcast.

The amplifier and system boxes, as seen in Fig. 8, can be bolted together or kept separate. The amplifier box has four antenna inputs, four amplified outputs, and a 3.3V power input. The main system box has four amplified signal inputs, 3.3V power output for the amplifier box, a main power input, a "heartbeat" pulse output, an external trigger input that is useful during tests, and output ports for Iridium and WiFi communications. The completed station boxes are roughly one cubic foot in size and set up very rapidly in the field.

Figure 9 shows the 2012 station electronics, consisting of four daughter-cards (one per-channel) and a motherboard, comprising the entire station except for the RF amplifiers and the two communications modules. A block diagram of the system is shown in Fig. 10.

Each daughter-card contains a 1.92 G-samples/s synchronous switched capacitor array analog sampling and digitization chip (the "ATWD"), a bias-tee module that adapts the DC offset of the incoming signal level to maximize the dynamic range of the ATWD, mode switches, power conditioning, DACs for threshold range settings, and a field-programmable gate array (FPGA) that aids in operating the ATWD and allows cards to operate as stand-alone devices if desired. Although the ATWD chips themselves include 128 10-bit analog to digital converters for fast parallel data conversion, a higher-resolution 12-bit ADC is included on each daughter-card for signal digitization.

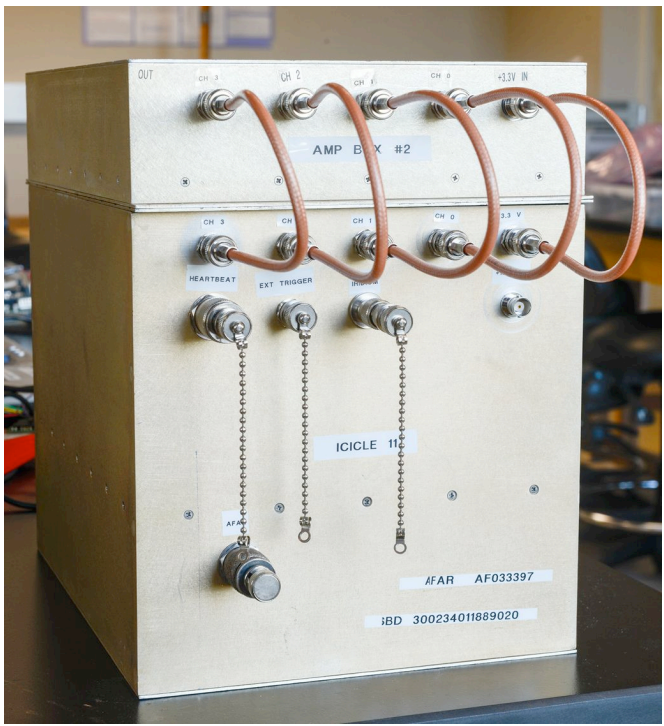

Fig. 8. An example of an HRA station's amplifier box (top portion) and main system box (bottom) containing all data acquisition, control and communications electronics. For scale, the width of the front of the box as shown is 9 inches, the depth is 12 inches, and the total height is 11.5 inches.

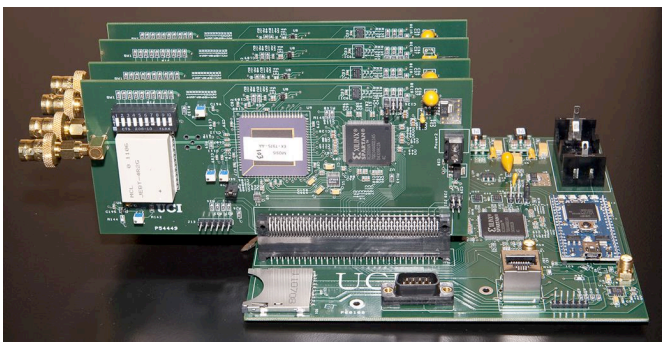

Fig. 9. The ARIANNA Hexagonal Array data acquisition electronics. It includes four 1.96 GHz data acquisition channels, a 100 MHz 32-bit CPU, communications channels for wireless and satellite short-burst message system, a 32 GB flash memory card holder for event data storage, power conditioning and control for all primary components, trigger I/O, "heartbeat" pulse generator, etc.

The system's motherboard contains all computing, communications interface hardware, data storage and power management circuitry necessary to run the station. It includes two-stage power regulation for the antenna amplifiers, a power I/O terminal block, solid state relays for peripheral power control, voltage and power monitoring circuitry, daughter-card power regulation and control, power regulation for the embedded CPU, a holder for a 1.5 Ah lithium battery backup for the real-time clock, a 100 MHz 32-bit ARM Cortex M3 micro-controller, an external trigger input, an FPGA programming port, an Ethernet port used for WiFi communications, an RS-232 port used for Iridium satellite messaging, a 32 G-Byte SDHC flash memory card slot, four daughter-card slots, an FPGA for fast system functions, and an output for the production of a fast "heartbeat" pulse.

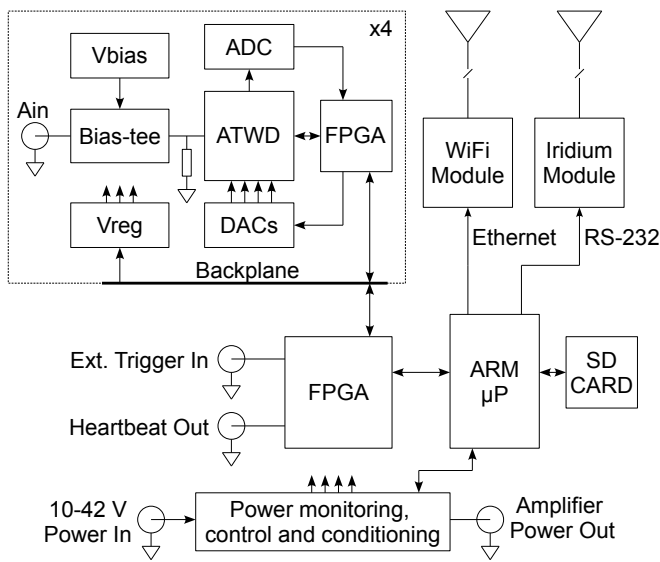

Fig. 10: Simplified block diagram of the ARIANNA system hardware.

## V. WAVEFORM ACQUISITION, TRIGGERING AND DIGITIZATION

Triggering and waveform capture is performed by a custom CMOS IC ("Advanced Transient Waveform Digitizer" or ATWD [19-22]) running at 1.92 G-samples/s and with ~11.5 bits of dynamic range [23, 24]. A block diagram of the chip's internals is shown in Fig. 11. The chip incorporates real-time pattern-matching trigger functionality that allows, for example, the detection of a bipolar waveform of a certain magnitude and frequency range. A prompt trigger is produced within about 15 ns of the targeted waveform's arrival.

### A. Sample rate and analog bandwidth.

The ATWD uses a synchronous sample clocking scheme that leads to high sample-to-sample timing uniformity. For convenience, it is driven by a low-speed external clock, which is boosted by a factor of 32 by an on-chip phase-locked loop system and then doubled via interleaving (using both clock phases) by an additional factor of two. The ARIANNA systems thus operates with a 30 MHz reference clock and achieves a net 1.92 GHz sample rate. By observing a test clock output from the ATWD chips with a histogramming period/frequency counter, the timing uniformity of this system has been measured to be ~1 ppm, RMS.

The analog bandwidth of the ATWD sampling and digitization system is an important figure of merit. With a 1.92 GHz nominal sample rate, the Nyquist-limited bandwidth would be 960 MHz, and ARIANNA's amplifiers are low-pass limited to approximately this frequency as well. Figure 12

*Preprint – October 26, 2014* 5

shows a plot of the frequency response of the data acquisition system as seen in Fig. 9 (excluding amplification). This plot was obtained by applying sine waves of a fixed amplitude of 500 mV but varying frequency (50 to 1,000 MHz in steps of 50 MHz). At each frequency, 1,000 waveforms were collected by the system and histogrammed. Two peaks are expected to appear in these histograms, corresponding to the high and low peaks of the sine waves as captured and digitized. The span of the peaks can thus be compared to the input sine wave, and as expected the magnitude of the output starts to drop as the bandwidth limitations of the system are reached. The frequency response of the entire system (excluding RF amplification) is seen to be relatively flat out to about 650 MHz, and its -3 dB frequency is about 860 MHz, near to the system's Nyquist limit of 960 MHz (1.92 GHz/2).

circuit. It also allows the comparators to be lower in power and slower, yet still, in effect, reach the full bandwidth of the ATWD's sampling system (i.e., ~860 MHz).

The basic high and low thresholds are set analogically via external DACs. However, as is the nature of all such electronic circuits, each comparator has a certain random input offset, and hence the ATWD chips include internal digital to analog conversion on a per-comparator basis to null these offsets for higher uniformity in triggering performance. To first order, the offsets are a form of "fixed pattern noise" and hence calibrations generally need to be done only once.

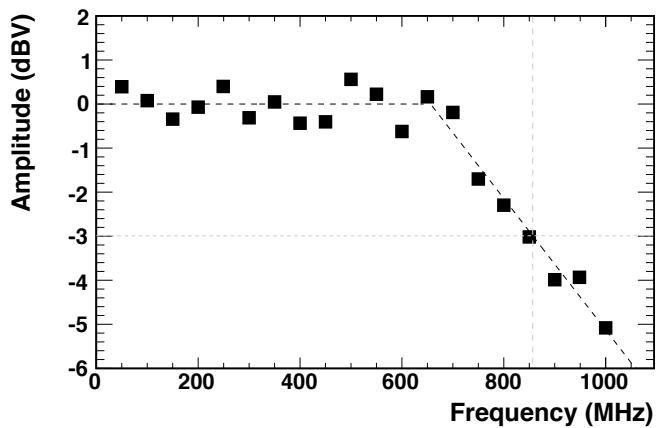

Fig. 12: Data acquisition system bandwidth measurements from digitized data, with the slope representing a fit to the higher-frequency data. The bandwidth is flat to ~650 MHz, and the -3dB bandwidth is ~850 MHz.

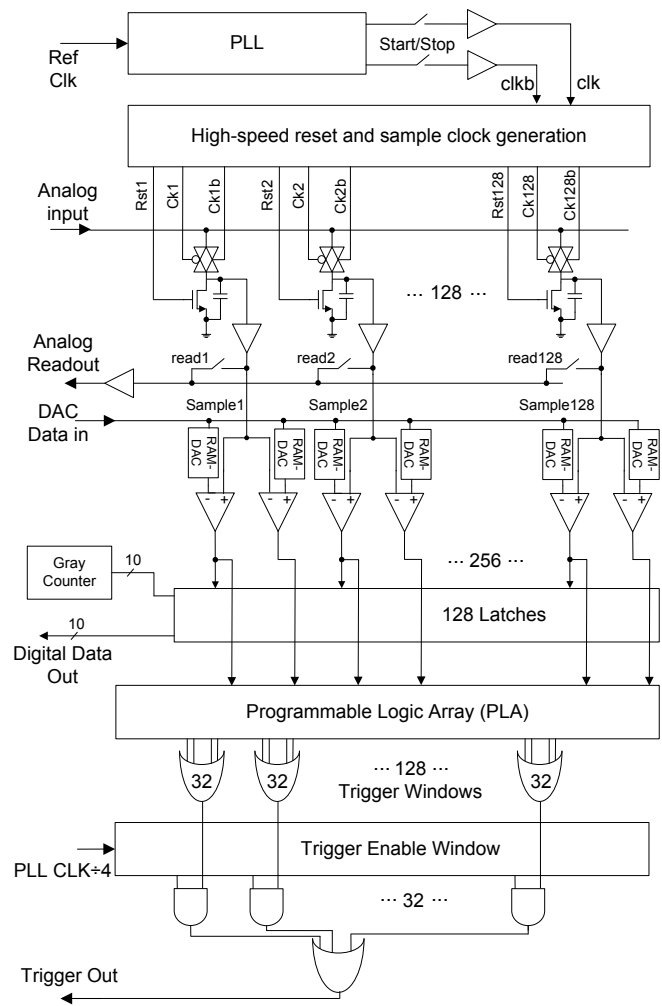

Fig. 11. A block diagram of the 2 GHz synchronous ATWD integrated circuit, showing sampling, comparison, and programmable trigger logic.

*A. Trigger thresholds and calibration.*

The ATWD chips perform dual (high and low) threshold triggering in real-time using a unique pattern-searching capability that observes the sampled signals rather than the input signal directly. This post-sample comparison does away with the need to split the input signal to a separate trigger

Figure 13 shows an example distribution of the offsets from one set of 128 comparator trigger thresholds (all "high" thresholds of a chip) before and after calibration. The "pulse height" axis represents the magnitude of a unipolar pulse at the channel's AC-coupled analog input that is narrow enough for its peak to be fully contained in one sample (the specific height is somewhat arbitrary, it's the trigger's response to varying heights as the pulse crosses a desired threshold that is of interest). These pulses, produced at 1 kHz, arrive asynchronously with respect to the ATWD's 1.92 GHz sample clock, and hence can arrive at any comparator's sample and hold. With ideal (zero) offsets, the transfer function between pulse height and trigger rate would be a step function from 0 Hz to 1 kHz at a single pulse height. However, in a realistic circuit, differing comparator input offsets lead to curves as seen in the figure. Nulling of these input offsets in this example is found to reduce variation in trigger thresholds from a sigma of 13.5 mV to a sigma of 3.6 mV. The latter number includes the noise of the signal generator itself, and yet is still less than a fifth of the sigma of the amplified thermal output noise from the amplifiers (~22 mV). Since such fixed pattern noise sources are independent and add only in quadrature to thermal noise, variations in trigger thresholds after calibration (in this case) results in only a ~2% net increase in noise in the trigger. The trigger offset nulling DAC values are stored on each daughter-card's FPGA's non-volatile memory, and are loaded into the ATWDs upon a command to cycle the data acquisition power.

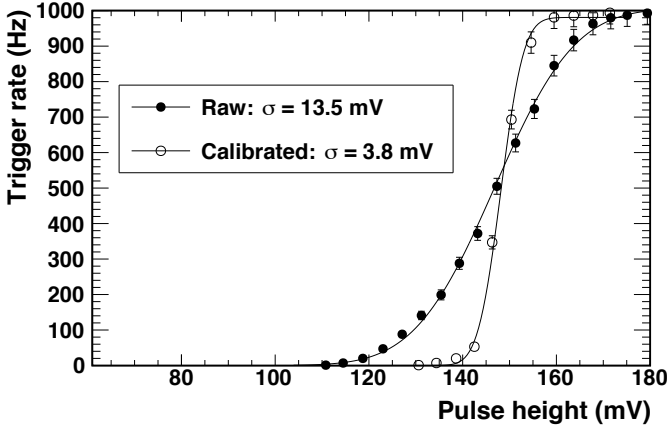 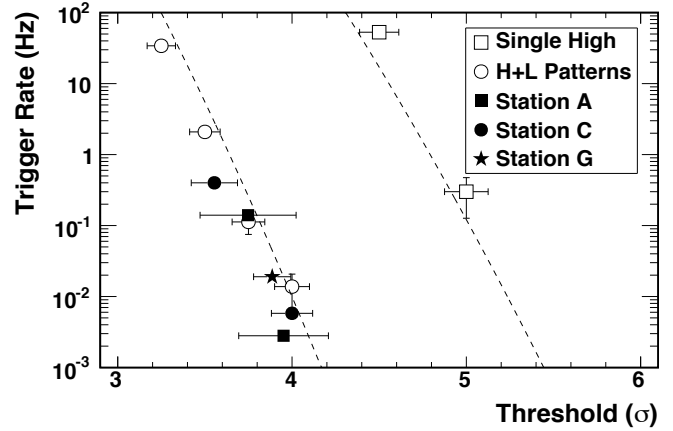

Fig. 13: Example trigger calibration for a single threshold showing trigger rate as a function of pulse height. The curve after calibration shows the variation in thresholds dropping by a factor of 3.6 to a sigma of 3.8 mV.

Fig. 14: Laboratory and *in-situ* measurements of trigger thresholds vs. trigger rates. The "Single High" laboratory measurements represent trigger rates for any crossing above a high threshold. The "H+L Patterns" represent laboratory measurements for a H and L trigger combination coincident within 4 ns. The "Station A, C and G" data points were from measurements made from three different stations, remotely collected during 2014 using the same "H+L Patterns" trigger criteria.

### B. Trigger rate control.

The ATWD has pattern-matching trigger capabilities that aids in trigger rate control [25]. Up to 72 patterns can be loaded into each chip. Each pattern may be a combination of input signal conditions, namely $H$ – the signal must be above a high threshold, $L$ – the signal must be below a separate low threshold, $N$ – neither above nor below (i.e., between the two threshold levels), or $X$ – don't care (does not veto triggers regardless of the signal level). Each pattern consists of 8 such conditions, representing 8 consecutive samples. A trigger pattern of *HXXXLXXX*, for example, looks for a bipolar signal in which a pair of high and low comparator values are about 2 ns apart (at 1.92 GHz, each sample is 0.52 ns apart).

ARIANNA further employs a second-level trigger that can require a coincidence between a combination of individual channel's triggers, with a programmable level of majority imposed (i.e., 1 or any 2, 3 or all 4 channels coincident within a certain time period). The combination of bipolar trigger patterns, programmable trigger thresholds, and second-level trigger majority logic can flexibly control trigger rates over many orders of magnitude. Furthermore, an advantage of a bipolar trigger over a simple unipolar threshold is that the former tends to stabilize trigger rates even when drifts, e.g., of the signal baseline, are occurring.

Figure 14 shows laboratory tests of the trigger rate vs. threshold while comparing two different patterns, plus *in-situ* measurements from the prototype HRA data. The threshold is expressed in terms of the amplifier noise sigma. The lines represent theoretical estimates of the expected rates. The "Single High Only" points denote trigger rates when a pattern of *HXXXXXXX* is used. This pattern will trigger on any over-threshold sample and is one of the most liberal patterns that can be used. The "H+L" patterns trigger on any signal that passes both the high and a low thresholds over a span of time ranging (technically) from 1.56-3.65 ns, i.e., the same range depicted in Fig. 13. This set of patterns is considerably more restrictive than the single threshold case yet more realistic for neutrino signatures. The resulting trigger rate drops by over 5 orders of magnitude for the same threshold values.

The *in-situ* measurements are from field data collected after two calibrations made during remote operation in early 2014. These calibrations and measurements were made with the same 5-pattern trigger criteria used in the "H+L" laboratory measurements. Note that all of the data shown in Fig. 17 are also using a "majority-2" criteria, namely that at least 2 channels must pass the individual channel's trigger criteria within a set period of time (in this case, ~64 ns).

The stability of the trigger under changing environmental conditions is also important. The ability to trigger on coincidences of High and Low thresholds has demonstrated significant gains in trigger stability across a wide range of temperatures. For example, a baseline drift due to a temperature change (e.g., due to the temperature coefficient of a voltage regulator) may cause the High threshold to increase in trigger frequency, while the Low threshold would decrease, countering each other and substantially moderating any net change in trigger rates.

Figure 15 shows trigger rates in a laboratory test of temperature stability. Two sets of data are shown, one with a High threshold only, and one with an equivalent High and Low coincidence required (over the space of 4 ns). For a single threshold (i.e., High only), a change such as a baseline drift of just a few mV will cause a significant change in trigger rates, and indeed the figure shows about two orders of magnitude change over approximately 15 degrees C, with at least another two orders of magnitude projected down to -30C (measurements were rate-limited to ~10 Hz, higher than expected *in-situ* rates). By contrast, using an equivalent High and Low coincidence results in about one order of magnitude change in trigger rates over the entire expected temperature range once buried in the snow of 0 to -30 C.

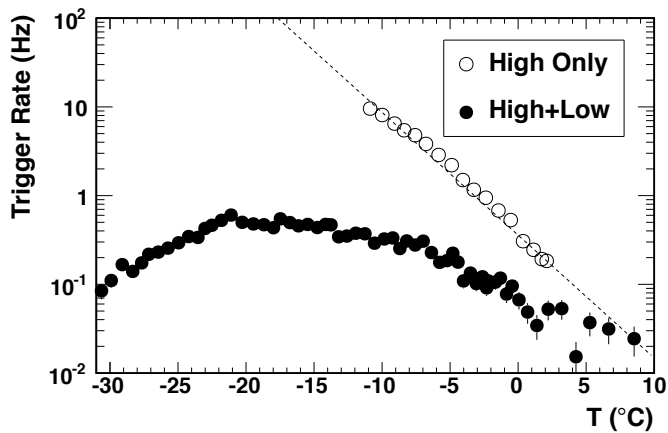

Fig. 15: Thermal trigger rates vs. temperature for a single high threshold trigger (open circles), and for a trigger that requires passing a high and a low threshold within 4 ns (filled circles).

A simple automatic threshold monitoring and adjustment system will eventually be put in place in ARIANNA's system software. However, ARIANNA's experience is that remotely-performed threshold changes need only be made a few times a year to remain within the system's range of operation and memory capacity. ARIANNA's end goal may be to maintain rates such that all data can be retrieved by Iridium, e.g., rates in the mHz regime, in order to reduce or eliminate any dependence on the high-speed WiFi link, and indeed ~2 mHz rates have been demonstrated in practice.

## VI. Power Systems

Given the ARIANNA site's isolation, and the 1 km distance between stations, each station is fully autonomous, including by producing their own power. For the 2012 deployment, the power system for each station consisted of three solar panels, experimental use of a wind-turbine, and lithium-iron-phosphate storage batteries.

### A. Power tower and solar panels.

Each station deployed in 2012 and prior years have been equipped with solar panels, an experimental wind turbine, and a battery for power buffering. The HRA stations deployed in 2012 used standard commercially-available radio tower components that were taller and quicker to assemble than the prior custom-made solution, and which integrated both solar and wind power. Each tower was 16 feet in height excluding the wind-turbine extension, and were tied-down by three steel cables connected to wooden anchors buried in the snow. Constructed almost completely of aluminum, the tower assemblies including solar and wind power systems were light enough to be raised manually by one individual.

Solar panels perform well in the Antarctic environment due to the high reflectivity of the snow. For the 2012 deployment, the ARIANNA power towers employed three solar panels in a triangular configuration. A primary 100W panel was oriented north, and provided more than sufficient power to run the station and maintain a peak battery charging state for nearly as long as the sun remains up. Two secondary 30W panels were mounted on the other two faces of the triangular tower for supplementary power when the sun is behind the main panel. During the summer, the solar panels provide enough power that the stations run continuously and exclusively on solar power nearly 100% of the time, even during periods of extensive cloud cover.

### B. Batteries at cold temperatures.

ARIANNA stations include batteries to store power for use during overcast days and weeks while the sun is rising and setting. $LiFePO_4$ batteries were selected based on this technology's high physical and chemical stability and safety, and after ARIANNA's experimental evidence of performance at cold temperatures. Each of the 3 HRA stations deployed thus far incorporated 2 $LiFePO_4$ batteries of 112 Ah nominal capacity when rated at room temperature (224 Ah total). These were configured in an automobile starting-battery form-factor (Braille Battery Co. model OSGC-12112iB). The batteries include integrated charge controllers which disconnect the batteries when fully charged (e.g., during summer when solar power is plentiful) and when the batteries are nearly depleted, to prevent damage from over-charging and over-discharging. Disconnects upon reaching full charge occurred transparently and did not cause any disruption of the station electronics. Disconnects when fully discharged, e.g. after the final setting of the sun for the year, caused the station to power down in an orderly fashion. Autonomous power-up has also been orderly.

ARIANNA conducted laboratory tests at -30C (previously measured to be the lowest expected winter temperature when buried in the snow) and demonstrated that the selected batteries retained about 70% of their nominal storage capacity when charged and discharged at these temperatures. At -30C, they were capable of accepting a charging current of at least 7A (ARIANNA's expected maximum), and easily provided the expected maximum discharge current consumed by the station electronics of 1A. Figure 16 displays an example charging and discharging profile of a single 112 Ah (nominal) battery at -30C. Starting from empty and at -30C, it required ~89 Ah of charge to reach a full state, at which point the charge controller disconnected the battery. From this state, discharged at 1 A, the battery delivered ~79 Ah of charge until it disconnected. Minor discontinuities in terminal voltages were seen at some points during transitions between a normal and cautionary state indicated by an LED on the battery housing that is driven by the battery's internal charge controller, presumably due to internal switching or rebalancing of cells.

Using ARIANNA's expected "worst case" usage profile, these results indicate a useful storage capacity of ~70% of one battery's nominal rating at -30C, and an efficiency of ~89%. With two batteries in parallel (as in the stations deployed in 2012), being charged and discharged at half of these rates per battery, slightly better performance may be expected. The net power available from two batteries stored, charged and discharged at -30C is thus at least 158 Ah, or enough to run the station by itself at full power for at least one week, and in low-power modes (e.g., with reduced data acquisition and communications duty cycles) for up to one month.

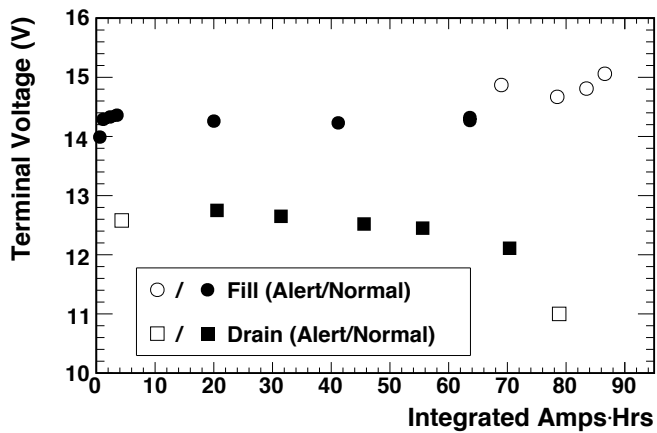

Fig. 16: Terminal voltage of a single battery charging at 7A and discharging at 1A, both at -30C. The "alert" vs. "normal" measurements denote when an LED on the battery housing indicated a nearly full or nearly empty status.

The batteries for the ARIANNA stations as deployed were contained in insulated enclosures so that any self-heating from charging and discharging them (i.e., from inefficiency losses) may warm the batteries, and to help stabilize any diurnal temperature changes. All connections between the power tower, the battery box and the station box were via bayonet connectors and hence were very fast and easy to complete while wearing gloves in the field.

*C. Wind power.*

Even at ~7W average power consumption, it has not been considered practical to power the stations by battery alone during the winter. Therefore, ARIANNA has experimented with a number of wind turbines. For the 2012 Austral summer deployment, each of the four stations (the three HRA stations plus the earlier prototype) were equipped with 150 watt maximum wind turbines (Primus Wind Power Air 40). As a precaution, the turbines were disassembled and their bearings re-packed with aircraft-grade grease rated to -70C. The Air-40 model uses glass-reinforced nylon blades which, in all stations, survived a year of operation without any issues. Unfortunately, the vertical rotation collar failed on one turbine, leaving it unbalanced and unable to transmit power. A second turbine failed when an internal mechanical part seized. Evidence points to both of these failures occurring during a single powerful storm. The third and fourth wind turbines remained functional.

During the 2013 servicing mission, the wind turbines were removed from all three HRA stations, which have hence-forth operated on solar power and batteries only. The Site D prototype station maintained its turbine for continued experimental use, and was reconfigured as an environmental monitoring station including air speed and temperature measurements. Figure 17 shows this station's data, supplemented with wind speed measurements made at Scott Base later in the season. Wind speeds at the ARIANNA site have been found to be sufficient for significant up-time during winter months, motivating continued interest in experimenting with wind power generation.

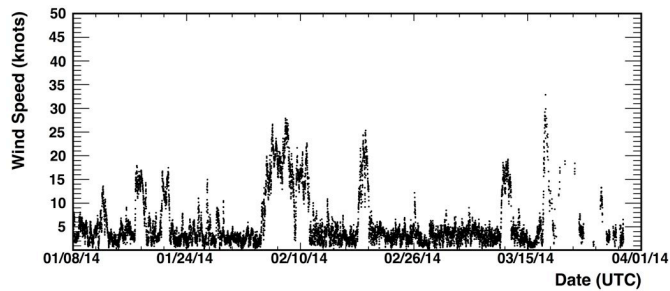

Fig. 17: Wind speed plot from Station 2's anemometer, supplemented at later dates with data from measurements made at Scott Base.

## VII. Monitoring, Control and Data Collection

The HRA stations are designed to operate autonomously, with remote monitoring, control and data collection possible by two redundant communications modalities – long distance wireless via a repeater located on Mt. Discovery and satellite short-burst messaging. Communications with each of the ARIANNA stations are handled by a custom software suite built in C++ and Python, via computing facilities at U.C. Irvine (UCI). The Python code works with the Twisted framework to handle TCP communications ("WiFi") and email communications (Iridium Short Burst Data messaging) to and from the stations. Multiple stations can and do communicate concurrently.

*A. Communications Overview*

The long-range wireless system allows fast and efficient retrieval of all station data, as well as control over each station, including the timing and duration of data acquisition and communications windows, control over which major subsystems are powered, and even the capability of loading new software for the stations' microcontrollers. For robustness, the station's wireless communications are mesh-connected, in that every station can act as a repeater for each other, and each can separately reach Mt. Discovery. Communications thus takes place through the "best" path, either directly from a station to McMurdo, or possibly first hopping through a different station that has a stronger signal. In particular, the early-prototype Site D station is currently set up as an always-on repeater.

As an alternative to the high-speed long-range wireless communications, each HRA station is equipped with an Iridium satellite short-burst data (SBD) messaging system. This provides functionality similar to that of a mobile-phone's text messaging system, with messages received by the station consisting of 270 bytes and sent messages containing 340 bytes. The SBD messaging system has been found to be very reliable; more so than the previous use of Iridium modem technology, which was prone to dropping connections. Although messages are short, they are densely encoded, and each transmits or receives a significant amount of control and monitoring data. Virtually every function available by WiFi is available by SBD. However, due to limited bandwidth, it is primarily used for control and monitoring, and to retrieve samples of data events. In this, it is important, as the ARIANNA stations have demonstrated longer calendar-time operational duration than the Mt. Discovery repeater, and so

Iridium is the only method of communicating with the stations during portions of the year. Finally, the Iridium receiver is used to synchronize each station's real-time clock to a highly-precise time received from satellites.

### B. System Software and Operation

Each HRA station's system software runs on an NXP LPC1768 embedded microcontroller using a 96 MHz ARM Cortex-M3 core with 32 kB of on-chip RAM and 512 kB of on-chip flash memory. Acquired data is stored on a 32 GB Compact Flash memory card, which is capacious enough to hold a year's worth of data or more even at the highest expected rates.

The system software is programmed in C/C++ without the benefits or overhead of a real-time operating system. The software breaks system operation into two major modes or "windows," namely communications and data taking. Generally, these alternate; when communicating, data taking is suppressed and powered-down, and during data taking, communications systems are powered down. Although it is possible to run the hardware associated with these modes simultaneously, these windows are kept separate because powering-up communications breaks RF-silence and may cause the collection of corrupted data. A communications window always precedes a data taking window to provide an opportunity to set parameters such as thresholds, window durations, etc.

At the start of a communications window, the system is normally set to attempt to use WiFi (the faster of the two) to send a status message and then waits for a response. If there is none (e.g., because the Mt. Discovery repeater is down), it defaults to SBD communications and tries again. If there is still no response, it will revert back to WiFi and make a programmable number of repeated tries. With no response in a certain number of tries, the system maintains its previously-programmed procedures, performs a data taking window, and tries to communicate again later. If communications are established, a control message is sent stating that there are no new commands (proceed as before), or else new commands and configurations can be sent to the station.

Because both WiFi and SBD communications are supported, effort has gone into creating a consistent, compact protocol that works well with both. Communications are in units termed "frames." Each frame has a 5-byte header that specifies the frame type and the size of its payload. A station can send data as an individual event (especially useful for sending a sample event via SBD) or entire data files.

Configuration commands sent to a station during a communications window include parameters such as trigger threshold levels, file and event compression parameters, and communications parameters such as timeout values for communications windows in case two-way communications are not established, the time between communications windows (equivalent to the duration of the data-taking windows), and what data to transmit during communications windows.

During data taking windows the system can perform data taking functions such as collecting "thermally" triggered events, periodic forced triggers in which the system takes an event unbiased by the trigger circuitry, and "heartbeat" events, in which the station generates an RF pulse itself and collects the resulting event. Data files collected during these windows include unique event numbers, time-stamp information, voltage readings, losslessly-compressed ADC values, bits confirming the type of trigger that resulted in the event (e.g., thermal, forced, etc.), and a 32-bit CRC value to aid in confirming data integrity.

The systems includes several features intended to enhance robustness, with particular attention to preventing a system from finding itself in some erroneous state whereby it may lose its ability to communicate, etc. These include a hardware-level "watchdog" timer that will completely reboot the system unless the watchdog is reset at least every 20 minutes (i.e., if the station becomes locked out of normal operation). The system will also completely reboot if it fails to achieve confirmed communications for 5 communications windows in a row. Furthermore, received control parameters are not allowed to fall outside of reasonable ranges to prevent user errors from accidentally disabling the stations. For example, the communications window duration is not allowed to be set to be less than 10 minutes. Finally, it is possible to remotely upload a new software revision to the station, which, if it passes a CRC check, etc., will take over. Thus far, there has not been cause to use this feature.

Since there are likely to be periods during which power conservation becomes important, it is possible individually control which of the major peripherals (amplifiers, data acquisition, WiFi, and Iridium SBD) are on or off during the communications and data-taking windows. For example, WiFi consumes substantially more power than Iridium SBD, and although it is much faster, the lower-power SBD system can be used exclusively when power savings becomes important. Finally, the systems can be placed in a strict power-savings mode, in which all data taking is powered down, and communications windows can be less frequent, etc. This mode can be entered automatically by a station when the battery voltage drops below a specified value. Hysteresis is implemented with a second value that prevents the station from dropping into and out of this mode too quickly. A very low-power mode gives operators the ability to maintain control when battery power is low.

### VIII. SYSTEM PERFORMANCE

The performance of the first three HRA systems have been extensively studied [26, 27]. This section describes the performance of the power systems, trigger rate performance and stability, noise performance, radio-pulse reflection studies, correlations to neutrino templates, station timing resolution, and event reconstruction resolution.

### A. Power systems performance.

As an example of the power system's performance, Fig. 18 shows voltage readings for an example station (Site A) during about 14 months of operation, from the time it was turned on in late November 2012 until late December, 2013, when it was disconnected for servicing, and subsequently through March 31, 2014. As evidenced by voltages in the 17-24V range (regions "A" and "E" in Fig. 23), the solar panels provided nearly all power during the Austral summer. In an interesting observation, the output voltage of the solar panels climbed during colder months, presumably due to lower levels of recombination and dark current in the solar cells, perhaps combined with the more direct angle of attack of the sun upon the panels. Battery power is seen supplementing the station's operation in voltage ranges of ~12 to ~14V (e.g., region "C"). Wind power was observed via voltages between ~14 and ~17V (region "B") to be frequently and strongly supplementing solar power from early February until mid-March of 2012, at which point the wind turbine evidently failed during a storm. Beyond this point, solar panels continued to provide significant power, and the batteries were observed to be fully charged during days out until mid-April, even at a time when the sun reached only about 2 degrees maximum height and only about one week before the last sunset on April 24, 2013. After the last sunset (first vertical line in Fig. 18), the station was alternately directed between normal and lower-powered modes in order to prolong testing of the station, e.g. of temperature effects, etc. During this time (region "D"), the station subsisted on battery power only.

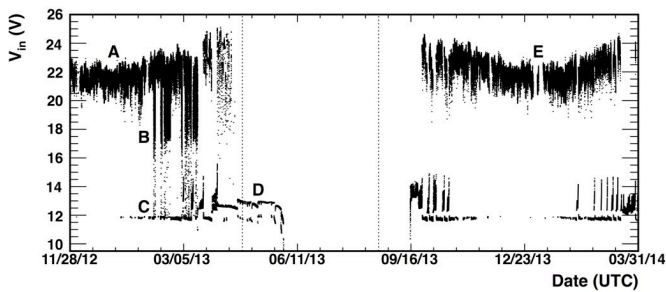

Fig. 18: Example station's power supply voltage vs. time. Periods of operation primarily on solar power ("A"), wind power ("B"), battery backup ("C") and solely on battery power ("D") are indicated. Period "E" shows interrupted power during station servicing, and also demonstrates a relationship between seasonal changes and solar power efficiency ("E" was mid-summer). The vertical lines indicate the last and first days of sun. The last day of operation was May 30, 2013 PST. The station made first contact again on September 12, 2013 PST and realized 100% up-time within four days later. Operation with 100% up-time was achieved during ~70% of a year.

In late May 2013, the battery's measured voltage began to decline precipitously, indicating that its reserve of power was close to exhaustion. Therefore, on May 28, 2013, the station was placed back in a full-power continuous data-taking and communications mode in order to test whether it would shut down in an orderly fashion and reboot autonomously from that state upon the return of the sun. The station thus shut down May 30, 2013, 36 days after the last sunset.

The first autonomous communication of the next spring occurred on September 12, 2013, about 3 weeks after the first sunrise (August 19, 2013, indicated by the second vertical line in Fig. 18). This was a day on which the sun had reached a maximum height of 8 degrees. On September 16, 2013, the station began uninterrupted operation until it was serviced in late December 2013. The station thus maintained 256 days of operation out of 365, or 70% of the year while including the use of power savings modes. When run at full power continuously, at least 58% of a year has been achieved.

### B. Trigger rates vs. temperature and wind.

Figure 19 shows an example station's (Site A) thermal-triggered event rate from January 2, 2014 and March 13, 2014. The amplifier's gain has been noted to rise slightly as temperature drops, leading to increased thermal trigger rates. Once the stations are covered in snow, diurnal temperature changes have been found to be less significant than seasonal changes. Since re-commissioning in January of 2014, the station's thresholds have been remotely adjusted twice, as noted by the two downward arrows in Fig. 20. All stations behaved similarly and required only the same two adjustments.

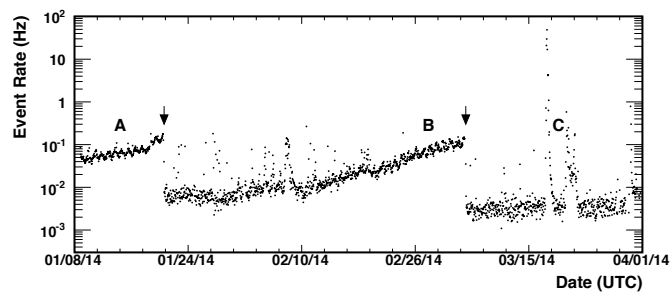

Fig. 19: Total event rates (triggered events only) vs. time for Site A from January 8 through March 31, 2014. During periods A and B, slight diurnal rates changes are visible, as is a gradual increase in rates related to a drop in temperature. Two adjustments in rates were made, indicated by the two downward arrows, on or about 1/23 and 3/06. During period C, a powerful storm swept through the area, and an increase in rates was noted.

A partial correlation between storms and/or wind velocity and event rates has been observed. In Fig. 20, the period "C," for example, shows an increase in rates during a storm. The cause and nature of the excess events is being studied, but a few comments can be made: Elevated event rates have been found to be correlated between stations. Generally, only wind speeds above ~20 knots have resulted in elevated event rates, but not all such periods of higher wind speeds have resulted in higher rates. Most of these temporary increases have had negligible impact on event collection efficiency, i.e., less than seasonal temperature variations. No additional noise has been found in forced (unbiased) events during storms, and so there is no evidence that increased trigger rates are due to any gradual, consistent change in the level of noise. Rather, these noise events appeared to be sparse and random. Only one few-hour-long instance (to the left of "C" in Fig. 19) resulted in excess event rates that significantly impacted dead-time, increasing it by ~1% over that limited time period. Analysis of the excess triggered events has concluded that they do not resemble expected neutrino events, and that these excess events can be removed from the data with high efficiency, as discussed in the next section.

## C. Thermal noise measurements.

Fig. 20 shows an example plot of recorded noise sigma in mV vs. time for the 2014 Station A, channel 2 data set, binned into 1-day intervals. The "Forced" time series consists of all data from "unbiased" events taken at periodic intervals without the involvement of the station's trigger system. This data is highly Gaussian and essentially displays the channel's thermal noise (average $\sigma$=17.6 mV for all forced triggers). A slight rise in noise vs calendar time is due to the slowly cooling temperatures, which has been found to increase the amplifier's gain and hence the level of recorded noise.

The "$\eta$>3" time-series, by contrast, contains all "thermally" triggered measurements (those acquired due to the system's trigger), but excluding events that have been identified with brief periods in which the station's amplifiers have displayed a sympathetic oscillation between channels (this problem has been rectified by a revised amplifier design as discussed in Section IX). The amplitudes measured by each sample in the triggered data are also substantially normally distributed, although amplitudes in at the trigger threshold values occur with a higher probability as expected. It is noted that fluctuations in triggered-event noise levels rise modestly above the unbiased event noise levels over the same storm or high-wind periods as seen in Fig. 17 and concomitant with the event rate increases seen in Fig. 19.

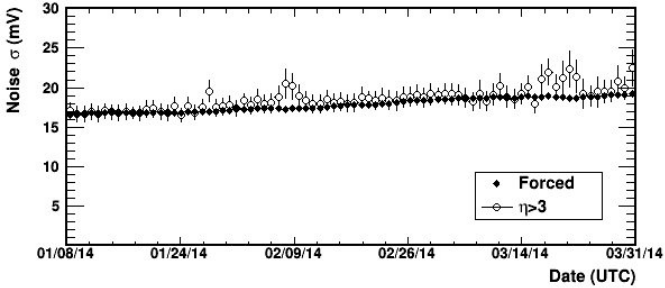

Fig. 20: Noise sigma for channel 2 at Site A between January 8, 2014 and March 31, 2014, binned into 1-day periods. The "Forced" data points are unbiased by the system's trigger and reflect highly-Gaussian thermal noise (average of 17.6 mV). The $\eta$>3 data reflects data collected due to the station's trigger system. Aside from a bias that the trigger imposes, this data is also substantially Gaussian, although episodes of greater noise are seen that are correlated with periods of storms including high winds.

## D. 2014 data-set correlation distributions.

Data taken between January 8, 2014 and March 31, 2014 has been studied in an exploratory search for neutrino-like signals [28]. An expected neutrino signal has been generated from the time dependent electric field at the neutrino interaction vertex, propagated through a model of the ice and convolved with measured antenna and amplifier response functions. The neutrino signals are determined as a function of two space angles defining the orientation of the incident electric field relative to the antenna, as well as the angle between the antenna and the Cherenkov cone. The resulting time dependent neutrino waveform "templates" (e.g., Fig. 21) can then be compared to recorded data by computing its maximum correlation value with each antenna waveform.

Prior to reconstruction of the event direction and polarization, waveforms from all four channels, including both the recorded waveform and its inverse (it is not *a priori* obvious which face of the antenna is presented to the incoming radio wave, hence whether the initial pulse would be positive or negative), for a total of 8 waveforms per station, are compared to a single reference template corresponding to 30° in the E and H-planes. The best correlation between any of these 8 signals and the reference template is designated as $\chi$. Figure 23 shows values of $\chi$ in the Station A data set for all events ("All Data") in its light-gray area, including a total of 203,562 events. In this analysis, which is described in much greater detail in [28], it is required that a neutrino candidate have a $\chi$>0.81.

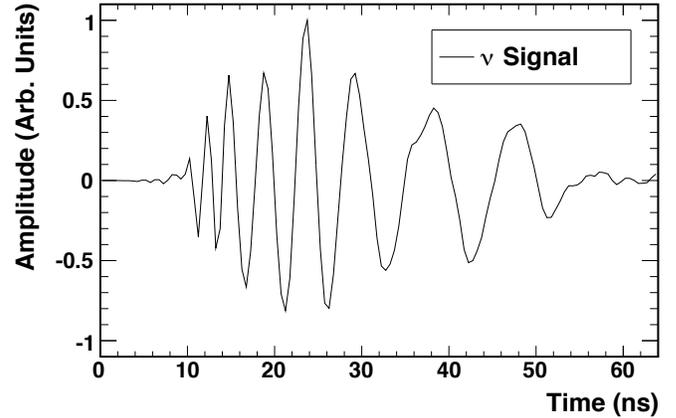

Fig. 21: Example neutrino signal template (40 degrees off-axis in the E-plane) including ice propagation, antenna, amplifier response, but excluding thermal noise, and sampled at 1.92 GHz (Y-axis units are arbitrary).

The majority of triggered events are purely random in nature (i.e., thermal noise). These are identified by an autocorrelation function whose results are noted to have a perfect correlation at zero time offset. Non-thermal-noise events are taken to be those for which the minimum autocorrelation function $\alpha$ is below -0.45 on any antenna. These remaining non-thermal events are shown in medium-gray in Fig. 22 ("$\alpha$<-0.45"), and comprise 25% of the "All Data" set. This cut preserves 99.5% of neutrino candidates.

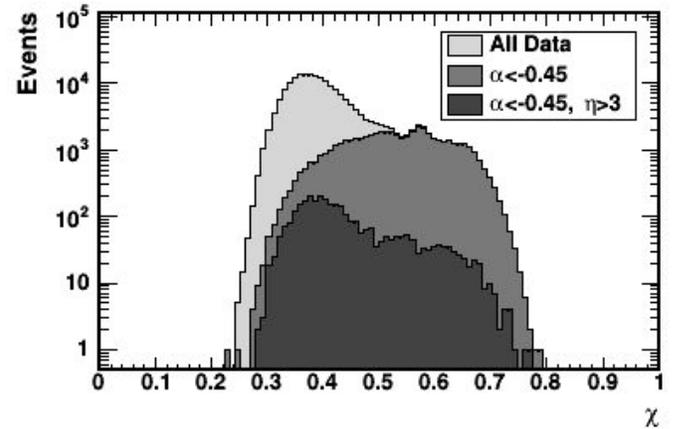

Fig. 22: Distribution of $\chi$ in the 2014 data set for Station A, channel 2, for all data, after the $\alpha$<-0.45 cut, and after both the $\alpha$ cut and the "$\eta$>3" cut.

The "η>3" cut mentioned in the previous sub-section is then made in addition to the "α<-0.45" cut. This, again, is intended to remove a small subset of events that contain sympathetic amplifier oscillations between channels. To pass this cut, it is required that the frequency spectrum of a neutrino candidate have more than 3 frequency bins (η>3) at or above 50% of the magnitude of the maximum bin – that is, that the candidate contains more than essentially the single-frequency oscillation that is seen in misbehaving amplifiers. This cut removes 85% of the events remaining after the α<-0.45 cut and preserves 97% of neutrino candidates. Long term, this cut may be unnecessary given amplifier stability improvements discussed in Section IX. The combination of the α and η cuts is seen in dark gray in Fig. 22, resulting in 3,159 remaining events (~1.5% of the full data set) and preserving 90% of expected cosmogenic neutrino events.

### E. Radio frequency reflection studies.

Radio-frequency reflection studies on one of the Site G HRA station have been performed. These involved delivering a fast electrical pulse, generated by a Pockels Cell driver (Grant Applied Physics model HYPS) to a quad-ridged polarization horn antenna (Seavey Engineering Inc., now Antenna Research Associates; antenna custom-designed for the ANITA project and described in [29]). The antenna was placed face-down to the ice at various locations both near to and far from the station, as well as oriented in several polarizations relative to the receiving antennas. The transmitted RF pulse therefore passed down through the ice (~550 m thick at Site G), bounced off of the water-ice interface, and back up to the station. The station electronics includes an external trigger input that allows the capture of waveforms at precise times. Inserting a controlled delay between the generation of the RF pulse and the station trigger was thus used to trigger the station's data acquisition at the time of arrival of the reflected pulse.

During the 2013-2014 service mission, a comparison of reflected waveforms was made between those collected by an ARIANNA station's electronics and equivalent waveforms using same ARIANNA channel's antenna and amplifier but captured by an oscilloscope (Agilent model DSO 7104B; 1 GHz bandwidth, 5 G-samples/s acquisition). As examples, two plots are shown from the same location (Station G), with the horn antenna located for a straight down-and-up reflection. The first comparison plot, Fig. 23, shows the station's channel 2's response to the reflected pulse (antenna oriented with parallel polarization to the transmitted pulse) superimposed on an equivalent pulse's response as recorded by the oscilloscope. Adjustments of the station's response to the vertical scale were made solely according to the station's calibration for gain. No adjustments to the oscilloscope's response was made.

Figure 24 compares a pulse received at the station's channel 1, whose antenna is orthogonal to that of channel 2 and thus orthogonal to the polarization of the transmitted pulse. Channel 1's response is attenuated compared to channel 2's, consistent with the difference in orientation. The polarization is evidently substantially maintained even after the reflection and transmission through a total of ~1100 meters of ice.

It can be seen that the waveforms within Fig.'s 23 and 24 are well-matched within the limits of noise (~22 mV RMS for the amplified thermal noise). It's also important to note that the overlapping waveforms shown in these figures are from different transmitted pulses, since it was not possible to record the same reflections at the oscilloscope and station simultaneously while using the same antennas and amplifiers. The evident degree to which the separate waveforms overlap therefore also supports the expectation that radio pulses traveling along identical trajectories through the ice and reflecting from the same patch on the ice-water interface are consistent from event to event, limited only by thermal effects.

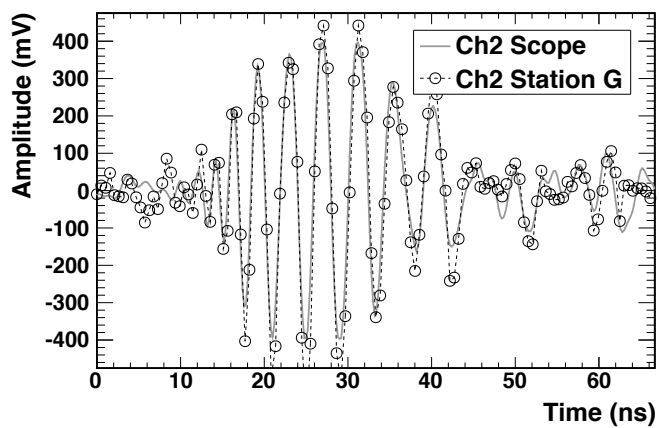

Fig. 23: An overlapping comparison of a representative antenna and amplifier response to separate but equivalent RF pulses reflected off of the bottom of the Ross Ice Shelf, as received by the Station G electronics and by a 1 GHz bandwidth oscilloscope. The polarization of the transmitted pulse was parallel to that of the receiving antenna.

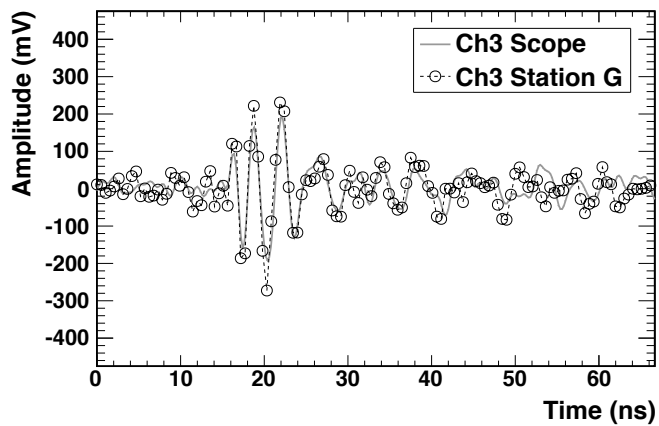

Fig. 24: An overlapping comparison of an antenna and amplifier response to separate but equivalent RF pulses reflected off of the bottom of the Ross Ice Shelf, as in Fig. 24. The polarization of the pulse was orthogonal to the receiving antenna and hence is attenuated.

### F. Station timing resolution.

Site G reflection studies, performed over a period of 24 hours for a variety of surface locations, have been was used to determine the station's timing resolution. For a given surface location, a reference event was arbitrarily selected to generate four $\Delta t_i$ values, where $\Delta t_i$ represents the time difference in the pulse arrival time between channel $i$ in the reference and current event. The time difference is taken to be that which maximizes the Pearson correlation between the waveforms on the $i^{th}$ channel in the reference and current event. This time difference may be non-zero due to jitter in the electronics used to generate the transmission pulse. However, all channels should have the same $\Delta t_i$ value, since jitter in the pulse transmission time should affect all channels equally. The difference in $\Delta t_i$ values between channels gives a measure of the readout timing resolution. Figure 25 shows the time difference $\Delta t = \Delta t_i - \Delta t_j$ for all six combinations of unique channel pairs $i$ and $j$, integrated over all events taken at all transmission locations. A net timing resolution of 0.049ns, obtained from a Gaussian fit to the peak, fully satisfies the experimental requirements of ARIANNA.

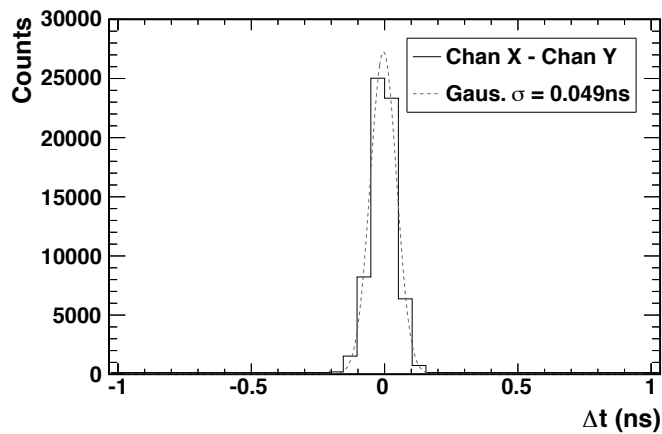

Fig. 25: Measured net timing resolution of the station at Site G, found via reflection studies initiated from a number of locations on the ice. The sigma of a fitted Gaussian is 0.049 ns.

### G. Angular resolution and event reconstruction.

Analysis of event reconstruction was performed using data taken in 2012 [30]. In brief, maximum cross-correlations were found between waveforms from all combinations of different channels. This leads to computed time differences between the channels and hence the angle at which a plane-wave is presumed to have struck the different antennas. The reconstructed angle at the station is then corrected for propagation through the firn layer (a layer of compacted snow from prior seasons) with a simple model of ice density as function of depth to produce a predicted signal-source location on the surface of the ice. The median value for the precision of the angular measurements for several different locations ranged between 0.14 to 0.17 degrees.

## IX. NEXT-GENERATION SYSTEM

A substantial redesign of the system electronics has been made, targeting deployment during the 2014-2015 Austral summer campaign. This has included the design of a new fast sampling chip, the development of a new single-board system to replace the motherboard/daughter-card system, and an updated amplifier design. The resulting hardware has improved electrical and physical robustness, better features and performance, uses substantially less power, is less costly, and is easier to calibrate. It maintains full "drop-in" compatibility with the installed HRA systems, yet facilitates easy scaling to 8-channel ARIANNA stations.

### A. Updated amplification.

An updated amplifier has been designed (Fig. 26), targeting deployment in 2014. It features enhanced stability, flatter frequency response and more symmetrical gain. To reduce system cost, it also eliminates the need for the external band-pass and limiting components seen in Fig. 5, with at most only a single attenuator needed to match the system board's input range.

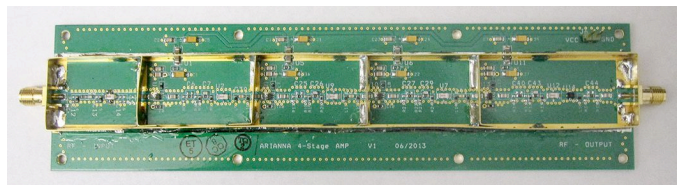

Fig. 26: Improved ARIANNA amplifier with cover shield removed.

### B. The SST, a new 2+ G-samples/s acquisition I.C.

A new signal acquisition integrated circuit has been designed and fabricated [31]. Containing 4 channels of 256 samples per channel, the "SST" Synchronous Sampling plus Triggering I.C. incorporates substantially the same trigger functionality as the ATWD system as described in Section VI, but in a greatly simplified, easier to use and lower-power form. The sampling is completely synchronous, using no PLL or any delay-based timing, and is simply driven by an external LVDS clock for extremely-high timing uniformity and stability. Because of its fully-synchronous design, the SST operates with clock rates spanning over 6 orders of magnitude, from <2 kHz to >2 GHz. Its leakage rate is so low (<200 mV/s) that operation down to <2 kHz is fully practical. Optimized design and packaging yielded a nearly-flat analog bandwidth to ~1.2 GHz using a standard 50-Ohm signal source and a -3 dB bandwidth of ~1.5 GHz. The use of an inexpensive 0.25 μm CMOS process allows a large input voltage range of 1.9V on a 2.5V supply. Table I summarizes some of the SST's main performance figures.

The SST includes a per-channel dual-threshold windowed coincidence trigger that operates with <1 mV RMS resolution and >600 MHz equivalent input bandwidth (e.g., it is sensitive to small-signal pulses down to 500 ps FWHM or better, with 0% to 100% triggers spanning less than 4mV in pulse height differences). An AND or an OR can be formed between comparators per channel over a window of ~3.5 ns or greater to form a bipolar trigger. For example, if set to 5ns, a bipolar signal of greater than 100 MHz, as is ARIANNA's specification, can be required in order to pass this first-level trigger. Output pins are available for each individual trigger comparator for easy calibration and rate monitoring or else, during typical operation, the AND of each channel's two comparators can be output in differential form. In AND mode, the SST stretches each trigger output to allow the simple formation of a second-level trigger that finds temporal coincidences between channels. The trigger outputs can be CMOS or will adapt to lower voltage levels (e.g., differential PECL) to help prevent noise coupling from the trigger outputs back to the analog inputs.

Table I: SST Figures of Merit

| Parameter | Value |
|---|---|
| Technology: | 0.25 μm CMOS |
| Supply voltage: | 2.5V |
| Number of channels: | 4 |
| Samples per channel: | 256 |
| Package size: | 8mm by 8mm |
| Number of package pins: | 56 |
| Input clock (ARIANNA): | 1 GHz LVDS |
| Sample rate (ARIANNA): | 2 GHz |
| Minimum sample rate: | < 2 kHz |
| Maximum sample rate: | > 2.5 GHz |
| Maximum power per channel: | 40 mW at 2 GHz |
| Analog input range: | 0-1.9V |
| Analog bandwidth: | > 1.5 GHz, -3dB |
| Dynamic range: | ~ 11.5 bits, RMS |
| Fixed pattern (pedestal) noise: | < 7 mV, RMS |
| Trigger comparators per channel: | 2 (high and low) |
| Trigger sensitivity: | < 1 mV, RMS |
| Trigger bandwidth: | > 600 MHz |
| Trigger functions per channel: | AND/OR, windowed |
| Trigger output modalities: | Differential/single-ended |
| Trigger output voltage: | 0.8, 1.2 or 2.5V CMOS |

The SST requires no programming, and only 3 active signals are required to operate it: Reset, Run/Stop, and Read-clock. The power consumption of the chip depends on the clock rate, the duty cycle of acquisition vs. digitization, and the bias on the comparators. When operating at the HRA's normal 2 G-samples/s acquisition speed, the worst-case power consumption is about 160 mW, or 40 mW per channel, with a more typical consumption of ~25 mW per channel.

### C. Next-generation system board.

A next-generation single-board data acquisition system board has been created, targeting use for the completion of the HRA in 2014. The new system, seen in Fig. 27, is designed for full physical and electrical drop-in compatibility with the HRA systems described in Section IV above. It includes one on-board 4-channel SST chip in place of four daughter-cards.

Power management has been improved, achieving higher input voltage tolerance (42V) along with the incorporation of on-board static discharge protection. The former was seen to be a potential necessity if wind power is reconsidered, while the latter was precautionary. A lower-power default turn-on state was also implemented, and attention was made to lowering parasitic power loss by the DC-DC converters and linear regulators that are on-board. The new system, and the SST in particular, has resulted in dramatic power reduction, from ~5.8W for the HRA system seen in Fig. 9 to ~1.7W for that in Fig. 27. Given the lower power consumption, the sensitivity and accuracy of the system board's voltage and current measurements was also enhanced. Finally, the board includes digital ambient temperature monitoring that is calibrated down to -55C.

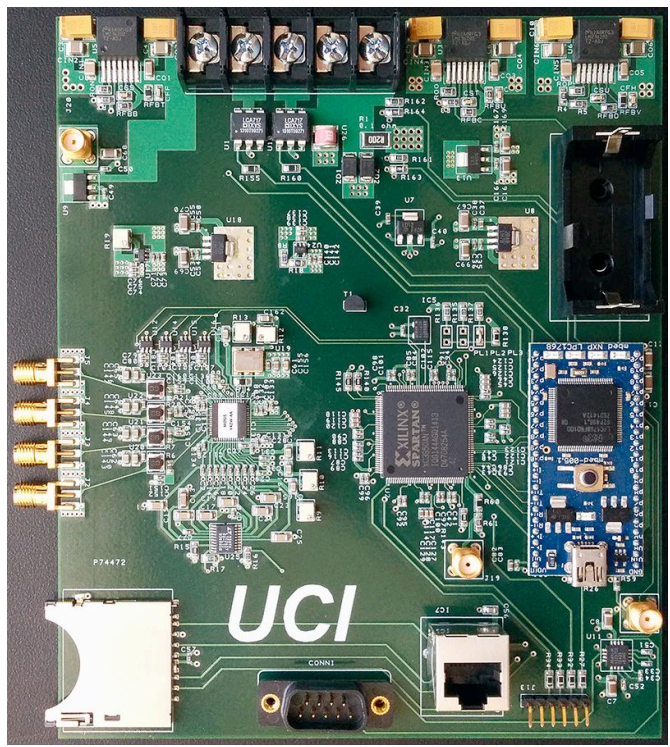

Fig. 27: The next-generation ARIANNA single-board data acquisition system. It includes one four-channel SST chip (left of center), improved power management and protection, on-board temperature monitoring, and has significantly-lower cost and calibration requirements. Average power consumption has been reduced from ~5.8W to ~1.7W.

The system firmware has also been significantly changed. In this version, all triggering and waveform digitization is managed independently by the board's FPGA. The FPGA forms second-level triggering (i.e., merging multiple channel's triggers), autonomously stops the SST, and digitizes its data, storing it in the FPGA's own block memory. Once digitization is completed, it delivers a flag to the system's microcontroller, indicating that an event has been taken and is available. This relieves the system's microcontroller from close, time-critical interaction, and permits higher levels of parallelism in a multi-SST system.

While the new system board was designed for electrical and physical compatibility with the first three HRA systems, it anticipates the creation of a compact, single-board, 8-channel version for full-scale ARIANNA stations. This would have the same or lower power consumption as the currently-deployed 4-channel systems, while offering a wider input range, longer record length, simpler trigger rate monitoring and calibration, easier fabrication and lower cost.

### D. Power systems changes for 2014.

For the 2014-2015 Austral summer completion of the Hexagonal Radio Array, the next four stations are currently planned for operation on solar power only. Therefore, the pyramidal upper tower segment and clearance necessary to accommodate a wind turbine as seen in Fig. 3 is not needed, and the upper tower will instead consist of a normal "straight" segment. This allows the solar panels to be mounted higher on the tower, allowing longer operation despite any snow accumulation. Furthermore, the deletion of wind power requires less battery capacity, since the primary purpose of larger batteries had been to buffer power between periods of higher wind when there is no sun available; a more stringent requirement than buffering cloudy periods. Hence, for 2014, only one battery of 112 Ah capacity (nominal) will be used per newly-installed station.

Since with solar power only there are no known or expected sources of anthropogenic RF noise remaining on the towers, the 2014 station's electronics will be placed at the foot of the power tower, with its signal antennas distributed about the tower, and the tower itself will be used to host the communications antennas. The resulting deletion of a separate communications mast further simplifies the system and reduces cost, deployment time and potential points of failure. Station electronics and batteries are planned to be co-located so that power used by the stations will help maintain an efficient battery temperature.

Experience and modeling have shown that the three-panel configuration used thus far (one north-facing 100W panel and two 30W panels) can be simplified to a single north-facing 100W panel without materially compromising useful powered life-time. The deletion of the wind turbine and the two auxiliary solar panels means that the tower's complexity and mass is reduced, and the power towers are thus simpler, quicker to set-up, lighter, and can be made taller if desired. It is expected that two 10-foot segments will be targeted for ARIANNA's full-scale deployment, gaining a further 4 feet of height with no added complexity.

Finally, instead of steel guying cable as used in the 2012 towers, Aramid cable will be used. This cable, marketed as a replacement for steel in these applications, including at low temperatures and under UV exposure, is much lighter and is essentially transparent to electromagnetic radiation. Hence, it can be used for guying near the RF antennas without influencing their performance.

SUMMARY AND CONCLUSIONS

The Hexagonal Radio Array is a pilot program for the ARIANNA project, intended to develop and demonstrate the technologies that the full project will rely on. Three HRA stations have been deployed, and each has gained over a year of operational exposure. The HRA systems have demonstrated low-power (~7W) and high-performance, including achieving 0.049 ns RMS timing resolution, ~2 mHz trigger rates at 4-sigma thresholds, 58-70% per-year operation on solar power and battery backup (depending on power and operation modes), angular reconstruction precision of 0.14 to 0.17 degrees, and with continuous remote monitoring, control and frequent full data transmission and control by long-distance wireless and satellite.

A simplification of several aspects of the HRA's systems hardware is being prepared with an aim of further reducing station power consumption, complexity, cost, calibration requirements and installation time per station. These include simplified power tower design and co-location of power, instrumentation and communications. An updated amplifier design with better stability, flatter frequency response, fewer external band-pass and limiting components has been fabricated. A new four-channel version of the ATWD waveform acquisition chip, the "SST," has also been designed, achieving twice the sample depth, 16 times lower power, wider input range, nearly twice the analog bandwidth, and much simpler operation. These simplifications and improvements are expected be incorporated in the last four HRA stations due to be deployed in the 2014-2015 Austral summer. They allow ARIANNA's full-scale data acquisition system to be created as a single board for reduced system cost, mechanical overhead and power consumption.

In conclusion, the technical goals that the pilot Hexagonal Radio Array sought have been substantially accomplished. The ARIANNA site has been proven to be virtually free from anthropogenic noise, autonomous operation with near-real-time full data retrieval has been demonstrated, and all critical system performance figures have been met. The 2014 hardware simplifications will allow full-scale ARIANNA systems to economically reach their goals, including the use of additional downward-facing antennas plus upward-facing antennas to co-locate the discrimination of cosmic rays and neutrino signatures.

ACKNOWLEDGMENTS

We wish to thank the staff of Antarctic Support Contractors, Lockheed, and the entire crew at McMurdo Station for excellent logistical support. We thank Wei Huang and Shiuh-hua Wood Chiang for their efforts in the design of the ATWD circuit, e.g. as depicted in Fig. 11. We also thank Prof. De Flaviis for the use of the Far Field Anechoic Chamber at U.C. Irvine.

This work was supported by generous funding from the Office of Polar Programs and the Physics Division of the US National Science Foundation, including via grant awards ANT-08339133, NSF-0970175, and NSF-1126672, and NSF-1126672, and by the Dept. of Physics and Astronomy, Uppsala University.

REFERENCES

[1] L. Gerhardt, et al,, "A Prototype Station for ARIANNA: A Detector for Cosmic Neutrinos," Nucl. Inst. Meth. A624, 2010, 85.
[2] S. W. Barwick, et al., "ARIANNA – A New Concept for High Energy Neutrino Detection," 32nd Intern. Cosmic Ray Conf., Beijing, 2011, pp 238-239.
[3] S. Klein, et al., "A radio detector array for cosmic neutrinos on the Ross Ice Shelf," IEEE Transactions on Nuclear Science, Vol. 60, No. 2, pp. 637-643, April 2013.
[4] S. A. Kleinfelder, et al., "Design and performance of the autonomous data acquisition system for the ARIANNA high energy neutrino experiment," IEEE Transactions on Nuclear Science, Vol. 60, Issue 2, pp 612-618, April 2013.
[5] G. A. Askaryan, JETP 14, 441 (1962); 21, 658 (1965).

[6] K. Dookayka, "Characterizing the Search for Ultra-High Energy Neutrinos with the ARIANNA Detector," Dissertation, University of California, Irvine, 2011.
[7] K. Dookayka, et al., "Characterizing the Search for UHE Neutrinos with the ARIANNA Detector," 32nd Intern. Cosmic Ray Conf., Beijing, 2011, 124.
[8] K. Greisen, "End to the Cosmic-Ray Spectrum?," Physical Review Letters 16 (17): 748–750, 1966.
[9] G. T. Zatsepin, V. A. Kuz'min, "Upper Limit of the Spectrum of Cosmic Rays". Journal of Experimental and Theoretical Physics Letters 4: 78–80, 1966.
[10] C. S. Neal, "The dynamics of the Ross Ice Shelf revealed by radio echo-sounding," Journal of Glaciology, 24, 295–307, 1979.
[11] S. Barwick, et al., "Performance of the ARIANNA Prototype Array," Proc. Int. Cosmic Ray Conf., Rio De Janeiro, July 2013.
[12] J. C. Hanson, et al., "Ross Ice Shelf Thickness, Radio-frequency Attenuation and Reflectivity: Implications for the ARIANNA UHE Neutrino Detector." Proc. Int. Cosmic Ray Conf., Beijing, China, 2011.
[13] T. Barrella, S. Barwick, D. Saltzberg, "Ross Ice Shelf in situ radio-frequency ice attenuation," J. Glaciology, Vol 57, No. 201, pp. 61-66, Feb. 2011.
[14] E. Andres, et al., "The AMANDA neutrino telescope: principle of operation and first results," Astroparticle Physics 13.1 (2000): 1-20.
[15] I. Kravchenko, et al. "Limits on the ultra-high energy electron neutrino flux from the RICE experiment," Astroparticle Physics 20.2 (2003): 195-213.
[16] J. Ahrens, et al., "Sensitivity of the IceCube detector to astrophysical sources of high energy muon neutrinos," Astroparticle Physics 20.5 (2004): 507-532.
[17] P. Allison, et al., "Design and Initial Performance of the 1351 Askaryan Radio Array Prototype EeV Neutrino Detector at the South Pole," Astropart. Phys. 35 (2012) 457–477.
[18] J. C. Hanson, et al., "Time-domain response of the ARIANNA detector," submitted to Astroparticle Physics.
[19] S. A. Kleinfelder, "A Multi-Gigahertz Analog Transient Waveform Recorder Integrated Circuit," 1992, Thesis, University of California, Berkeley.
[20] S. A. Kleinfelder, "Advanced transient waveform digitizers," Proceedings of the SPIE Particle Astrophysics Instrumentation, Vol. 4858, pp. 316-326, August, 2002.
[21] S. A. Kleinfelder, "A multi-GHz, multi-channel transient waveform digitization integrated circuit," Proceedings of the 2002 IEEE Nuclear Science Symposium, Orlando, FL, October 2002.
[22] S. A. Kleinfelder, "GHz waveform sampling and digitization circuit design and implementation," IEEE Transactions on Nuclear Science, Vol. 50, No. 4, pages 955-962, August 2003.
[23] W. Huang, S.W. Chiang, S. Kleinfelder, "Waveform Digitization with Programmable Windowed Real-Time Trigger Capability," Proceedings of the 2009 IEEE Nuclear Science Symposium, Orlando, FL, October 2009.
[24] S. A. Kleinfelder, et al., "Multi-GHz waveform sampling and digitization with real-time pattern-matching trigger generation," IEEE Transactions on Nuclear Science, Vol. 60, No. 5, pp. 3785-3792, October 2013.
[25] M. Roumi, "Advanced Pattern-Matching Trigger System for the ARIANNA High-Energy Neutrino Detector," Dissertation, U.C. Irvine, 2014.
[26] J. Tatar, "Performance of Sub-Array of ARIANNA Detector Stations during First Year of Operation," Dissertation, U.C. Irvine, 2013.
[27] J. C. Hanson, "The Performance and Initial Results of the ARIANNA Prototype," Dissertation, U.C. Irvine, 2013.
[28] S.W. Barwick, et. al., "A First Search for Cosmogenic Neutrinos with the ARIANNA Hexagonal Radio Array, submitted to Astroparticle Physics, 2014.
[29] P. W. Gorham et al., "The Antarctic Impulsive Transient Antenna Ultra-high Energy Neutrino Detector (ANITA): Design, performance, and sensitivity for 2006-2007 balloon flight," Astroparticle Physics, 32:10–41, 2009.
[30] C. Reed, et al., "Performance of the ARIANNA Neutrino Telescope Station," Proc. Int. Cosmic Ray Conf., Rio De Janeiro, July 2013.
[31] S. A. Kleinfelder, E. Chiem, T. Prakash, "The SST Fully-Synchronous Multi-GHz Analog Waveform Recorder with Nyquist-Rate Bandwidth and Flexible Trigger Capabilities," Conference Record of the 2003 IEEE Nuclear Science Symposium, 2014, in press.